\documentclass[longauth, letter]{aa} 
\usepackage{graphicx}
\usepackage{natbib}
\usepackage{amssymb}
\usepackage{amsmath}
\usepackage{url}
\usepackage[section]{placeins}
\usepackage{multirow}
\usepackage{txfonts}
\newcommand{\change}[1]{#1}          

\begin{document}
   \title{\textit{Herschel} observations of the hydroxyl radical (OH) \\in young stellar objects\thanks{\textit{Herschel} is an ESA space observatory with science instruments provided by European-led Principal Investigator consortia and with important participation from NASA.}}

\author{S.F.~Wampfler\inst{\ref{inst3}}
\and G.J.~Herczeg\inst{\ref{inst2}}
\and S.~Bruderer\inst{\ref{inst3}}
\and A.O.~Benz\inst{\ref{inst3}}
\and E.F.~van~Dishoeck\inst{\ref{inst2},\ref{inst1}}
\and L.E.~Kristensen\inst{\ref{inst1}}
\and R.~Visser\inst{\ref{inst1}}
\and S.D.~Doty\inst{\ref{inst19}}
\and M. Melchior\inst{\ref{inst3},\ref{inst46}}
\and T.A.~van~Kempen\inst{\ref{inst28}}
\and U.A.~Y{\i}ld{\i}z\inst{\ref{inst1}}
\and C.~Dedes\inst{\ref{inst3}}
\and J.R.~Goicoechea\inst{\ref{inst16}}
\and A.~Baudry\inst{\ref{inst6}}
\and G.~Melnick\inst{\ref{inst28}}
\and R.~Bachiller\inst{\ref{inst12}}
\and M.~Benedettini\inst{\ref{inst13}}
\and E.~Bergin\inst{\ref{inst14}}
\and P.~Bjerkeli\inst{\ref{inst9}}
\and G.A.~Blake\inst{\ref{inst15}}
\and S.~Bontemps\inst{\ref{inst6}}
\and J.~Braine\inst{\ref{inst6}}
\and P.~Caselli\inst{\ref{inst4},\ref{inst5}}
\and J.~Cernicharo\inst{\ref{inst16}}
\and C.~Codella\inst{\ref{inst5}}
\and F.~Daniel\inst{\ref{inst16}}
\and A.M.~di~Giorgio\inst{\ref{inst13}}
\and C.~Dominik\inst{\ref{inst17},\ref{inst18}}
\and P.~Encrenaz\inst{\ref{inst20}}
\and M.~Fich\inst{\ref{inst21}}
\and A.~Fuente\inst{\ref{inst22}}
\and T.~Giannini\inst{\ref{inst23}}
\and Th.~de~Graauw\inst{\ref{inst10}}
\and F.~Helmich\inst{\ref{inst10}}
\and F.~Herpin\inst{\ref{inst6}}
\and M.R.~Hogerheijde\inst{\ref{inst1}}
\and T.~Jacq\inst{\ref{inst6}}
\and D.~Johnstone\inst{\ref{inst7},\ref{inst8}}
\and J.K.~J{\o}rgensen\inst{\ref{inst24}}
\and B.~Larsson\inst{\ref{inst25}}
\and D.~Lis\inst{\ref{inst26}}
\and R.~Liseau\inst{\ref{inst9}}
\and M.~Marseille\inst{\ref{inst10}}
\and C.~M$^{\textrm c}$Coey\inst{\ref{inst21},\ref{inst27}}
\and D.~Neufeld\inst{\ref{inst29}}
\and B.~Nisini\inst{\ref{inst23}}
\and M.~Olberg\inst{\ref{inst9}}
\and B.~Parise\inst{\ref{inst30}}
\and J.C.~Pearson\inst{\ref{inst31}}
\and R.~Plume\inst{\ref{inst32}}
\and C.~Risacher\inst{\ref{inst10}}
\and J.~Santiago-Garc\'{i}a\inst{\ref{inst33}}
\and P.~Saraceno\inst{\ref{inst13}}
\and R.~Shipman\inst{\ref{inst10}}
\and M.~Tafalla\inst{\ref{inst12}}
\and F.F.S.~van der Tak\inst{\ref{inst10},\ref{inst11}}
\and F.~Wyrowski\inst{\ref{inst30}}
\and P.~Roelfsema\inst{\ref{inst10}}
\and W.~Jellema\inst{\ref{inst10}}
\and P.~Dieleman\inst{\ref{inst10}}
\and E.~Caux\inst{\ref{inst47},\ref{inst48}}
\and J.~Stutzki\inst{\ref{inst43}}
}

\institute{
Institute for Astronomy, ETH Zurich, 8093 Zurich, Switzerland\label{inst3}
\and
Max Planck Institut f\"{u}r Extraterrestrische Physik, Giessenbachstrasse 1, 85748 Garching, Germany\label{inst2}
\and
Leiden Observatory, Leiden University, PO Box 9513, 2300 RA Leiden, The Netherlands\label{inst1}
\and
Department of Physics and Astronomy, Denison University, Granville, OH, 43023, USA\label{inst19}
\and 
Institute of 4D Technologies, University of Applied Sciences NW, CH-5210 Windisch, Switzerland\label{inst46}
\and
Harvard-Smithsonian Center for Astrophysics, 60 Garden Street, MS 42, Cambridge, MA 02138, USA\label{inst28}
\and
Centro de Astrobiolog\'{\i}a. Departamento de Astrof\'{\i}sica. CSIC-INTA. Carretera de Ajalvir, Km 4, Torrej\'{o}n de Ardoz. 28850, Madrid, Spain.\label{inst16}
\and
Universit\'{e} de Bordeaux, Laboratoire d'Astrophysique de Bordeaux, France; CNRS/INSU, UMR 5804, Floirac, France\label{inst6}
\and
Observatorio Astron\'{o}mico Nacional (IGN), Calle Alfonso XII,3. 28014, Madrid, Spain\label{inst12}
\and
INAF - Istituto di Fisica dello Spazio Interplanetario, Area di Ricerca di Tor Vergata, via Fosso del Cavaliere 100, 00133 Roma, Italy\label{inst13}
\and
Department of Astronomy, The University of Michigan, 500 Church Street, Ann Arbor, MI 48109-1042, USA\label{inst14}
\and
Department of Radio and Space Science, Chalmers University of Technology, Onsala Space Observatory, 439 92 Onsala, Sweden\label{inst9}
\and
California Institute of Technology, Division of Geological and Planetary Sciences, MS 150-21, Pasadena, CA 91125, USA\label{inst15}
\and
School of Physics and Astronomy, University of Leeds, Leeds LS2 9JT, UK\label{inst4}
\and
INAF - Osservatorio Astrofisico di Arcetri, Largo E. Fermi 5, 50125 Firenze, Italy\label{inst5}
\and
Astronomical Institute Anton Pannekoek, University of Amsterdam, Kruislaan 403, 1098 SJ Amsterdam, The Netherlands\label{inst17}
\and
Department of Astrophysics/IMAPP, Radboud University Nijmegen, P.O. Box 9010, 6500 GL Nijmegen, The Netherlands\label{inst18}
\and
LERMA and UMR 8112 du CNRS, Observatoire de Paris, 61 Av. de l'Observatoire, 75014 Paris, France\label{inst20}
\and
University of Waterloo, Department of Physics and Astronomy, Waterloo, Ontario, Canada\label{inst21}
\and
Observatorio Astron\'{o}mico Nacional, Apartado 112, 28803 Alcal\'{a} de Henares, Spain\label{inst22}
\and
INAF - Osservatorio Astronomico di Roma, 00040 Monte Porzio catone, Italy\label{inst23}
\and
SRON Netherlands Institute for Space Research, PO Box 800, 9700 AV, Groningen, The Netherlands\label{inst10}
\and
National Research Council Canada, Herzberg Institute of Astrophysics, 5071 West Saanich Road, Victoria, BC V9E 2E7, Canada\label{inst7}
\and
Department of Physics and Astronomy, University of Victoria, Victoria, BC V8P 1A1, Canada\label{inst8}
\and
Centre for Star and Planet Formation, Natural History Museum of Denmark, University of Copenhagen,
{\O}ster Voldgade 5-7, DK-1350 Copenhagen K., Denmark\label{inst24}
\and
Department of Astronomy, Stockholm University, AlbaNova, 106 91 Stockholm, Sweden\label{inst25}
\and
California Institute of Technology, Cahill Center for Astronomy and Astrophysics, MS 301-17, Pasadena, CA 91125, USA\label{inst26}
\and
the University of Western Ontario, Department of Physics and Astronomy, London, Ontario, N6A 3K7, Canada\label{inst27}
\and
Department of Physics and Astronomy, Johns Hopkins University, 3400 North Charles Street, Baltimore, MD 21218, USA\label{inst29}
\and
Max-Planck-Institut f\"{u}r Radioastronomie, Auf dem H\"{u}gel 69, 53121 Bonn, Germany\label{inst30}
\and
Jet Propulsion Laboratory, California Institute of Technology, Pasadena, CA 91109, USA\label{inst31}
\and
Department of Physics and Astronomy, University of Calgary, Calgary, T2N 1N4, AB, Canada\label{inst32}
\and
Instituto de Radioastronom\'{i}a Milim\'{e}trica (IRAM), Avenida Divina Pastora 7, N\'{u}cleo Central, E-18012 Granada, Spain\label{inst33}
\and
Kapteyn Astronomical Institute, University of Groningen, PO Box 800, 9700 AV, Groningen, The Netherlands\label{inst11}
\and
Centre d'Etude Spatiale des Rayonnements, Universit\'e de Toulouse [UPS], 31062 Toulouse Cedex 9, France\label{inst47}
\and 
CNRS/INSU, UMR 5187, 9 avenue du Colonel Roche, 31028 Toulouse Cedex 4, France\label{inst48}
\and
KOSMA, I. Physik. Institut, Universit\"{a}t zu K\"{o}ln, Z\"{u}lpicher Str. 77, D 50937 K\"{o}ln, Germany\label{inst43}
}

\date{\today} \titlerunning{\textit{Herschel} hydroxyl observations of YSOs}


\def\placefigureHIFISpectrum{
\begin{figure}
 \centering
 \resizebox{6.3cm}{!}{\includegraphics[angle=270]{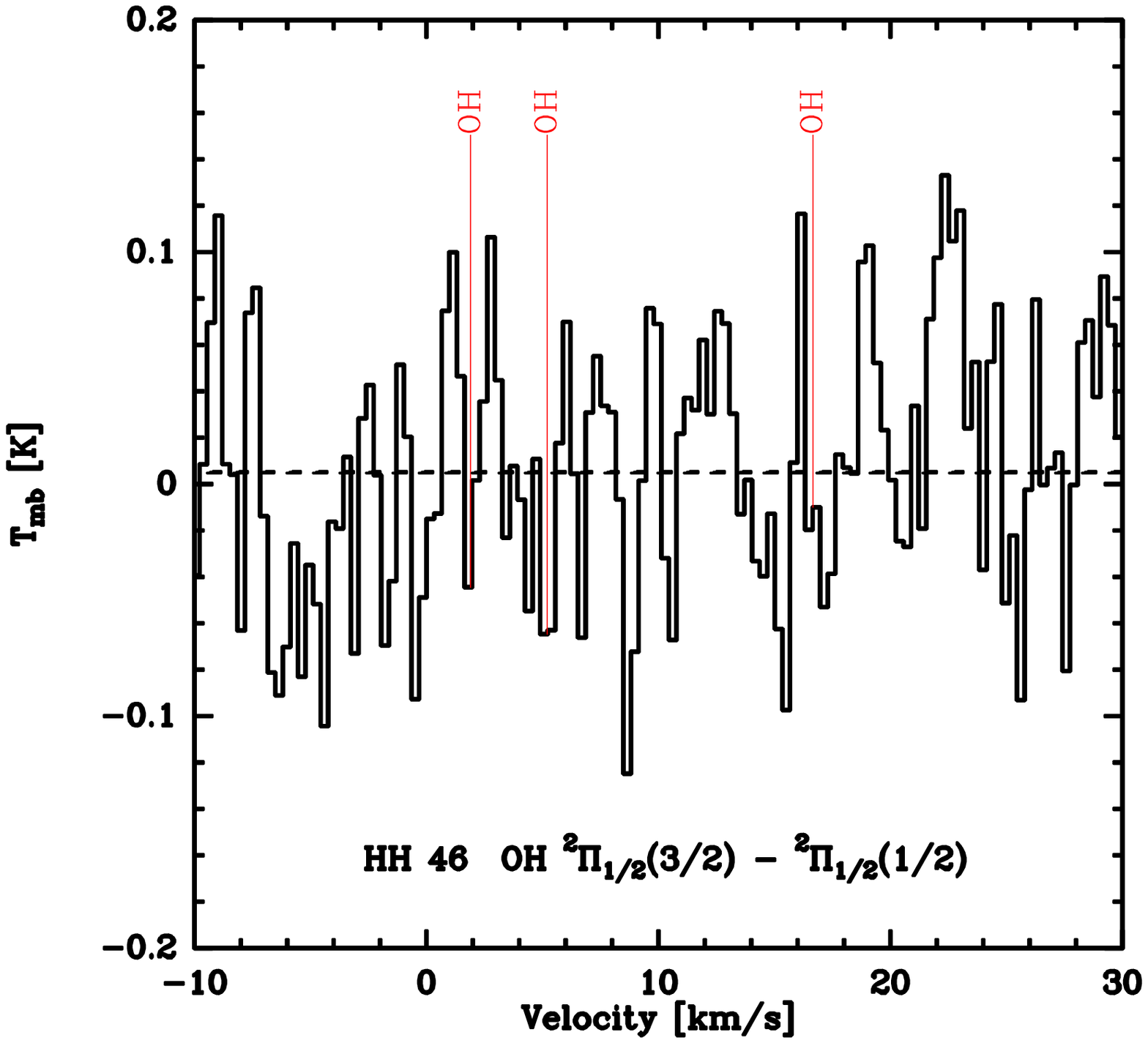}}
 \caption{HIFI spectrum at \change{163.12$~\mu$m (1837.8~GHz)} towards the low-mass YSO HH~46 ($v_{\mathrm{lsr}} = 5.2~\mathrm{km}~\mathrm{s}^{-1}$). OH lines were not detected \change{in 34~min of on source integration time (polarizations combined)}.}
 \label{fig:hh46_hifi}
\end{figure}
}

\def\placefigurePACSpectrum{
\begin{figure}
 \centering
 \resizebox{\hsize}{!}{\includegraphics[angle=0,bb=64 360 568 720]{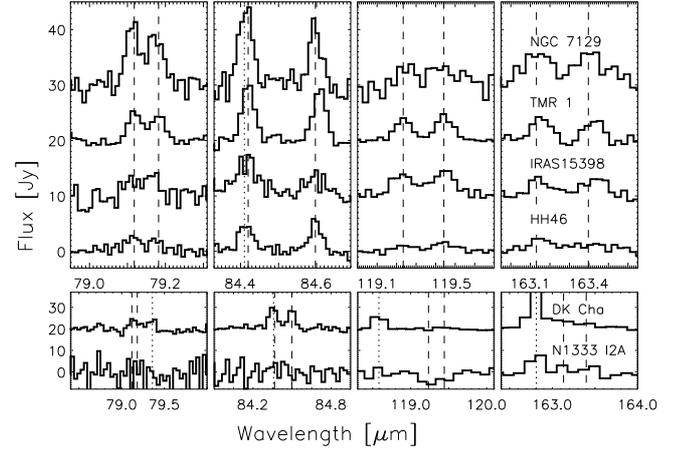}}
 \caption{PACS spectra of the observed OH doublets at 79, 84, 119, and 163$~\mu$m. Sources in the top panel were observed in line spectroscopy mode, \change{those} in the lower panel \change{in} range spectroscopy \change{mode} (different sampling). Dashed vertical lines indicate the OH frequencies, dotted lines show the position of CO lines.}
 \label{fig:pacs_spectra}
\end{figure}
}

\def\placefigureFluxEup{
\begin{figure}
 \centering
 \resizebox{0.92\hsize}{!}{\includegraphics[angle=0]{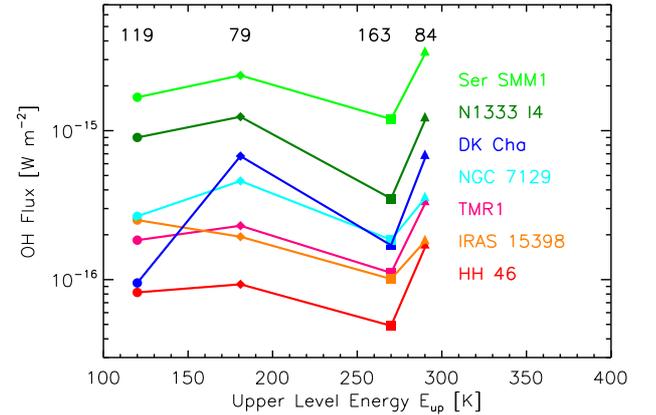}}
 \caption{Observed OH line fluxes from PACS plotted versus the upper level energy of the transition. Ser~SMM~1 and NGC~1333~IRAS~4 measured with ISO \change{are included for comparison}. Symbols correspond to different OH transitions: 79$~\mu$m as diamond, 84$~\mu$m as triangle, 119$~\mu$m as circle, and 163$~\mu$m as square. \change{Fluxes of the doublet components are summed. The $84.60~\mu$m flux multiplied by two is used for the $84~\mu$m doublet because the $84.42~\mu$m line is blended with CO(31-30).}}
 \label{fig:flux_vs_eup}
\end{figure}
}

\def\placefigureFluxLum{
\begin{figure}
 \centering
 \resizebox{0.92\hsize}{!}{\includegraphics[angle=0]{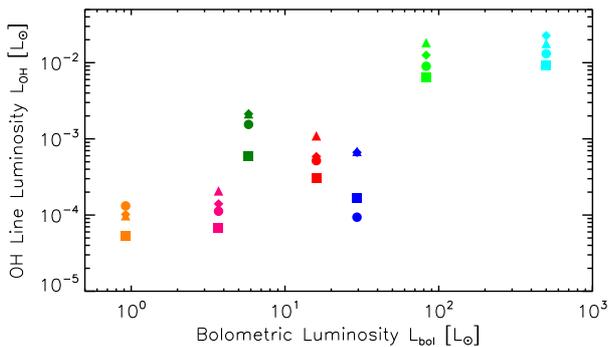}}
 \caption{\change{Dependence of OH line luminosity on the bolometric luminosity of the source}. Symbols and colors as in Fig. \ref{fig:flux_vs_eup}.}
 \label{fig:flux_lum}
\end{figure}
}

\def\placefigureFluxOxygen{
\begin{figure}
 \centering
 \resizebox{0.92\hsize}{!}{\includegraphics[angle=0]{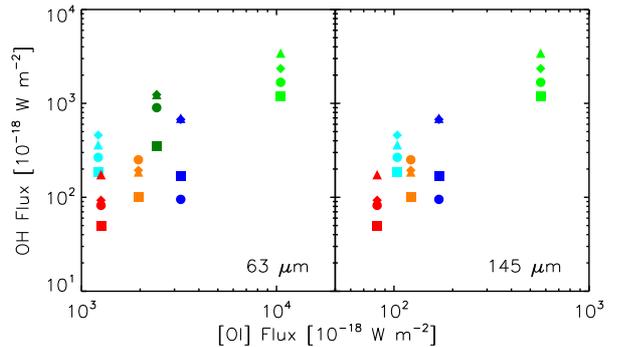}}
 \caption{Observed OH fluxes plotted against [\ion{O}{i}] 63$~\mu$m flux (left panel) and 145$~\mu$m flux (right panel). Symbols and colors as in Fig. \ref{fig:flux_vs_eup}.}
 \label{fig:flux_ox}
\end{figure}
}

\def\placefigureOHladder{
\begin{figure}[!h]
 \centering
 \resizebox{\hsize}{!}{\includegraphics{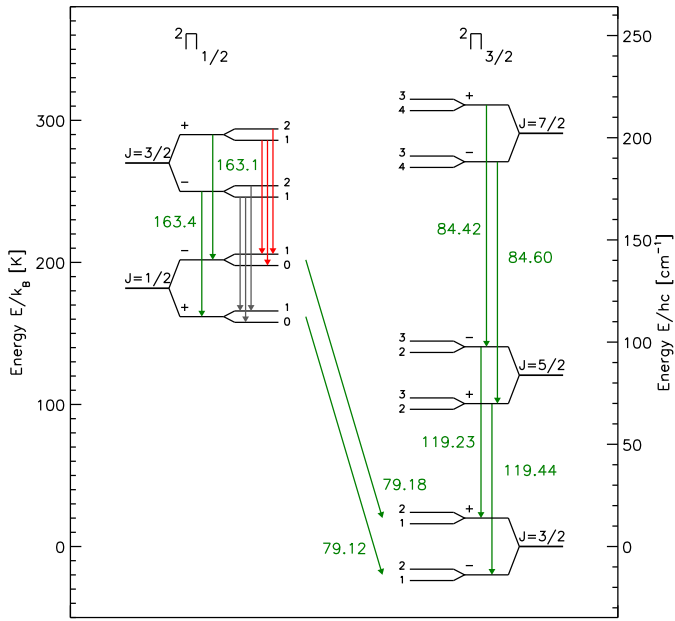}}
 \caption{Level diagram of the lowest excited states of OH up to \mbox{$E_{\mathrm{up}} \approx 300~\mathrm{K}$}. Splitting of the levels because of $\Lambda$-doubling and hyperfine structure is not to scale. Transitions observed with PACS are shown in green, the high-resolution observations of the hyperfine transitions carried out with HIFI in red.}
 \label{fig:oh_levels}
\end{figure}
}

\def\placefigureRatios{
\begin{figure}[!h]
 \centering
 \resizebox{\hsize}{!}{\includegraphics[angle=0]{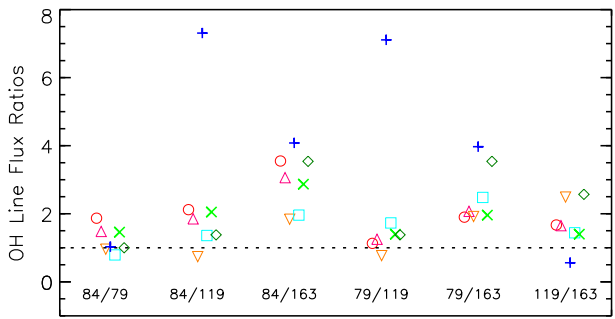}}
 \caption{Ratios of the observed OH fluxes. The numbers denote the corresponding wavelengths. \change{The symbols are: circle for HH~46, upward triangle for TMR~1, downward triangle for IRAS~15398-3359, squares for NGC~7129~FIRS~2, plus signs for DK~Cha, crosses for Ser~SMM1 and diamonds for NGC~1333~IRAS~4}. The color coding is the same as in Fig. \ref{fig:flux_vs_eup}.}
 \label{fig:oh_ratios}
\end{figure}
}


\def\placetableone{
\begin{table*}
\caption{OH \change{and $[$\ion{O}{i}$]$} line fluxes observed with PACS.}
\scriptsize
\centering
\begin{tabular}{lr@{\extracolsep{2pt}}rrrrrrrr}
\hline \hline
                               &         &   &  & \multicolumn{1}{c}{HH~46} & \multicolumn{1}{c}{TMR~1} & \multicolumn{1}{c}{IRAS~15398}
& \multicolumn{1}{c}{DK~Cha} & \multicolumn{1}{c}{NGC~7129} & \multicolumn{1}{c}{N~1333~I~2A}\\
\hline
Transition                     & \multicolumn{1}{c}{$\lambda$} & \multicolumn{1}{c}{$\nu$} & \multicolumn{1}{c}{$E_{\mathrm{up}}$} & \multicolumn{1}{c}{Flux}       & \multicolumn{1}{c}{Flux}  &  \multicolumn{1}{c}{Flux}  & \multicolumn{1}{c}{Flux} &  \multicolumn{1}{c}{Flux} & \multicolumn{1}{c}{Flux} \\
			      & \multicolumn{1}{c}{$[\mu\mathrm{m}]$} & \multicolumn{1}{c}{[GHz]}  & \multicolumn{1}{c}{[K]} &\multicolumn{1}{c}{$[10^{-18}~\mathrm{W}~\mathrm{m}^{-2}]$} &\multicolumn{1}{c}{$[10^{-18}~\mathrm{W}~\mathrm{m}^{-2}]$} & \multicolumn{1}{c}{$[10^{-18}~\mathrm{W}~\mathrm{m}^{-2}]$} & \multicolumn{1}{c}{$[10^{-18}~\mathrm{W}~\mathrm{m}^{-2}]$} & \multicolumn{1}{c}{$[10^{-18}~\mathrm{W}~\mathrm{m}^{-2}]$} & \multicolumn{1}{c}{$[10^{-18}~\mathrm{W}~\mathrm{m}^{-2}]$}\\
\hline
OH $\frac{1}{2},\frac{1}{2} - \frac{3}{2},\frac{3}{2}$  & 79.12 & 3789.3 & 181.9 &  55 $\pm$ \phantom{0}7 
										 & 128 $\pm$ 14 
                                                                                 & 194 $\pm$ \phantom{0}29 $^{\rm a}$           
										 & 360 $\pm$ \phantom{0}87 \phantom{$^{\rm a}$}             
										 & 229 $\pm$ \phantom{0}33
										 & $<$350  \\ 
OH $\frac{1}{2},\frac{1}{2} - \frac{3}{2},\frac{3}{2}$  & 79.18 & 3786.3 & 181.7 &  38 $\pm$ \phantom{0}5 
										 & 102 $\pm$ 14 
										 &  -$^{\rm a}$                           
										 & 315 $\pm$ \phantom{0}58 \phantom{$^{\rm a}$}             
										 & 230 $\pm$ \phantom{0}39
										 & -       \\ 
OH $\frac{3}{2},\frac{7}{2} - \frac{3}{2},\frac{5}{2}$  & 84.60 & 3543.8 & 290.5 &  87 $\pm$ \phantom{0}6 
										 & 170 $\pm$ 18 
										 & \phantom{0}93 $\pm$ \phantom{0}16 \phantom{$^{\rm a}$} 
										 & 347 $\pm$ \phantom{0}48 \phantom{$^{\rm a}$}             
										 & 181 $\pm$ \phantom{0}28
										 & $<$165  \\ 
OH $\frac{3}{2},\frac{5}{2} - \frac{3}{2},\frac{3}{2}$  &119.23 & 2514.3 & 120.7 &  38 $\pm$ \phantom{0}9 
										 &  83 $\pm$ 10 
										 & 121 $\pm$ \phantom{0}17 \phantom{$^{\rm a}$} 
										 &  95 $\pm$ \phantom{0}50 $^{\rm a}$   
										 & 132 $\pm$ \phantom{0}39 
										 & absorption    \\ 
OH $\frac{3}{2},\frac{5}{2} - \frac{3}{2},\frac{3}{2}$  &119.44 & 2510.0 & 120.5 &  44 $\pm$ \phantom{0}7 
										 & 101 $\pm$ 11 
										 & 130 $\pm$ \phantom{0}17 \phantom{$^{\rm a}$} 
										 &  -$^{\rm a}$                                      
										 & 134 $\pm$ \phantom{0}59
										 & absorption    \\ 
OH $\frac{1}{2},\frac{3}{2} - \frac{1}{2},\frac{1}{2}$  &163.12 & 1837.8 & 270.1 &  22 $\pm$ \phantom{0}4 
										 &  56 $\pm$ \phantom{0}8 
										 &  47 $\pm$ \phantom{00}9 \phantom{$^{\rm a}$} 
										 & 170 $\pm$ \phantom{0}60 $^{\rm a}$   
										 & 116 $\pm$ \phantom{0}45
										 & $<$100  \\ 
OH $\frac{1}{2},\frac{3}{2} - \frac{1}{2},\frac{1}{2}$  &163.40 & 1834.7 & 269.8 &$<$27\phantom{ $\pm$ 00} 
										 &  55 $\pm$ \phantom{0}7
										 &  54 $\pm$ \phantom{0}10 \phantom{$^{\rm a}$} 
										 &  -$^{\rm a}$                                      
										 &  69 $\pm$ \phantom{0}26 
										 & -       \\ 
$[$\ion{O}{i}$]$ $^3$P$_{1}$--$^3$P$_{2}$               & 63.18 & 4744.8 & 227.7 &1260 $\pm$ 54           
										 & -  
										 &1958 $\pm$ 197 \phantom{$^{\rm a}$}           
										 &3128 $\pm$ 333 \phantom{$^{\rm a}$} 
										 &1219 $\pm$ 130 
										 & $<$320  \\
$[$\ion{O}{i}$]$ $^3$P$_{0}$--$^3$P$_{2}$               &145.53 & 2060.1 & 326.6 &  82 $\pm$ \phantom{0}8 
										 & -  
										 & 122 $\pm$ \phantom{0}17 \phantom{$^{\rm a}$} 
										 & 170 $\pm$ \phantom{0}72 \phantom{$^{\rm a}$} 
										 & 104 $\pm$ \phantom{0}35
										 & $<$180  \\ 
\hline
\end{tabular}
\vspace{-4pt}
\begin{list}{}{}
\item[$^{\mathrm{a}}$] Doublet not resolved. Table lists the integrated flux over both components.
\end{list}
\label{tab:fluxes}
\end{table*}
}

\def\placetabletwo{
\begin{table*}[!H]
\caption{Source properties and observational details.}
\scriptsize
\centering
\begin{tabular}{l@{\extracolsep{2pt}}rrrl|l@{\extracolsep{2pt}}llll}
\hline \hline
Source & Distance & Luminosity    & Envelope Mass  & Type & RA      & Dec            & Obs. Date & Obs.id \\ 
       & [pc]     & [L$_{\odot}$] & [M$_{\odot}$]  &      & [h m s] & [${}^{\circ}~{\arcmin}~{\arcsec}$] &           &        \\
\hline
HH~46             & 450$^{\mathrm{a}}$ &  16\phantom{.00}$^{\mathrm{b}}$   &   5.1\phantom{0}$^{\mathrm{b}}$  & Class I        &
08:25:43.9  & $-$51:00:36\phantom{.0}  & 2009-10-26 & 1342186315 (PACS)\\
                  &     &       &        &                &            &                        & 2010-04-17 & 1342194783 (HIFI)\\
TMR~1             & 140$^{\mathrm{c}}$ &   3.7\phantom{0}$^{\mathrm{d}}$ &   0.12$^{\mathrm{e}}$ & Class I        &04:39:13.7  & +25:53:21\phantom{.0}  & 2010-03-29 & 1342192984 (PACS)\\
IRAS~15398-3359   & 130$^{\mathrm{f}}$ &   0.92$^{\mathrm{g}}$&   0.5\phantom{0}$^{\mathrm{h}}$  & Class I        &15:43:01.3  & $-$34:09:15\phantom{.0}  & 2010-02-27 & 1342191302 (PACS)\\
DK~Cha            & 178$^{\mathrm{i}}$ &  29.4\phantom{0}$^{\mathrm{j}}$ &   0.03$^{\mathrm{j}}$ & Herbig Ae      &12:53:17.2  & 
$-$77:07:10.6            & 2009-12-10 & 1342188039 (PACS)\\
                  &     &       &        &                &            &                        & 2009-12-10 & 1342188040 (PACS)\\
NGC~7129~FIRS~2   &1260$^{\mathrm{k}}$ & 500\phantom{.00}$^{\mathrm{l}}$   &  50\phantom{.00}$^{\mathrm{m}}$    & Intermed.-Mass &
21:43:01.7  & +66:03:23.6            & 2009-10-26 & 1342186321 (PACS)\\
NGC~1333~I~2A     & 235$^{\mathrm{n}}$ &  20\phantom{.00}$^{\mathrm{o}}$ &   1.0\phantom{0}$^{\mathrm{o}}$  & Class 0        &
03:28:55.6  & +31:14:37\phantom{.0}  & 2010-02-13 & 1342190686 (PACS)\\
                  &     &       &        &                &            &                        & 2010-02-24 & 1342191149 (PACS)\\
                  &     &       &        &                &            &                        & 2010-03-08 & 1342191773 (HIFI)\\  
\hline
Serpens~SMM1      & 415$^{\mathrm{p}}$ &  82.9\phantom{0}$^{\mathrm{q}}$ &   8.7\phantom{0}$^{\mathrm{q}}$  & Class 0        &            &                        & -          & ISO$^{\mathrm{r}}$              \\
NGC~1333~I~4      & 235$^{\mathrm{n}}$ &   5.8\phantom{0}$^{\mathrm{o}}$ &   4.5\phantom{0}$^{\mathrm{o}}$  & Class 0        &            &                        & -          & ISO$^{\mathrm{s}}$              \\
\hline
\end{tabular}
\begin{list}{}{}
\item[$^{\mathrm{a}}$] Heathcote, S., Morse, J. A., Hartigan, P., et al. 1996, AJ, 112, 1141
\item[$^{\mathrm{b}}$] van Kempen, T. A., van Dishoeck, E. F., G\"usten, R. et al. 2009, A\&A, 501, 633
\item[$^{\mathrm{c}}$] Motte, F., Andre, P., \& Neri, R. 1998, A\&A, 336, 150
\item[$^{\mathrm{d}}$] Ohashi, N., Hayashi, M., Kawabe, R. et al. 1996, ApJ, 466, 317
\item[$^{\mathrm{e}}$] J\o{}rgensen, J. K., Sch\"{o}ier, F.L., \& van Dishoeck, E.F. 2002, A\&A, 389, 908
\item[$^{\mathrm{f}}$] Murphy, D. C., Cohen, R., \& May, J. 1986, A\&A, 167, 234
\item[$^{\mathrm{g}}$] Froebrich, D. 2005, ApJS, 156, 169
\item[$^{\mathrm{h}}$] van Kempen, T. A., van Dishoeck, E. F., Hogerheijde, M. R. et al., 2009, A\&A, 508, 259
\item[$^{\mathrm{i}}$] Whittet, D. C. B., Prusti, T., Franco, G. A. P., et al. 1997, A\&A, 327, 1194
\item[$^{\mathrm{j}}$] van Kempen, T. A., Green, J. D., Evans, N. J. et al. A\&A, 2010, accepted
\item[$^{\mathrm{k}}$] Shevchenko, V.S. and Yakubov, S.D. 1989, SvA, 33, 370
\item[$^{\mathrm{l}}$] Johnstone, D., Fich, M., M$^{\textrm c}$Coey, C. et al. 2010, this volume
\item[$^{\mathrm{m}}$] Crimier, N., Ceccarelli, C., Alonso-Albi, T. et al. 2010, arxiv 1005.0947
\item[$^{\mathrm{n}}$] Hirota, T., Bushimata, T., Choi, Y. K., et al. 2008, PASJ, 60, 37
\item[$^{\mathrm{o}}$] J\o{}rgensen, J. K., van Dishoeck, E. F., Visser, R., et al. 2009, A\&A, 507, 861
\item[$^{\mathrm{p}}$] Dzib, S., Loinard, L., Mioduszewski, A.J., et al.  2010, ApJ, in press, arXiv1003.5900
\item[$^{\mathrm{q}}$] Hogerheijde, M. R., van Dishoeck, E. F., Salverda, J. M. et al., ApJ, 513, 350 ($L_{\mathrm{bol}}$ scaled to a distance of 415~pc)
\item[$^{\mathrm{r}}$] Larsson, B., Liseau, R. \& Men'shchikov, A.B. 2002, A\&A, 386, 1055 
\item[$^{\mathrm{s}}$] Giannini, T., Nisini, B. \& Lorenzetti, D. 2001, ApJ, 555, 40  
\end{list}
\label{tab:source_props}
\end{table*} 
}


\abstract
   {}
   {`Water in Star-forming regions with \textit{Herschel}' (WISH) is a \textit{Herschel} Key Program investigating the water chemistry in young stellar objects (YSOs) during protostellar evolution. Hydroxyl (OH) is one of the reactants in the chemical network most closely linked to \change{the formation and destruction of H$_2$O}. 
   \change{High-temperature ($T \gtrsim 250~\mathrm{K}$) chemistry connects OH and H$_2$O through the OH + H$_2$ $\Leftrightarrow$ H$_2$O + H reactions. Formation of \change{H$_2$O} from OH is efficient in the high-temperature regime found in shocks and the innermost part of protostellar envelopes. Moreover, in the presence of UV photons, OH can be produced from the photo-dissociation of H$_2$O through H$_2$O + $\gamma_{\mathrm{UV}}$ $\Rightarrow$ OH + H.}}
   {High-resolution spectroscopy of the \change{163.12$~\mu$m} triplet of OH towards HH~46 and NGC~1333~IRAS~2A was carried out with the Heterodyne Instrument for the Far Infrared (HIFI) \change{on board the \textit{Herschel Space Observatory}}. The low- and intermediate-mass protostars HH~46, TMR~1, IRAS~15398-3359, DK~Cha, NGC~7129~FIRS~2, and NGC~1333~IRAS~2A were observed with the Photodetector Array Camera and Spectrometer (PACS) \change{on \textit{Herschel}} in four transitions of OH and two [\ion{O}{i}] lines.}
   {The OH transitions at 79, 84, 119, and 163$~\mu$m and [\ion{O}{i}] emission at 63 and 145$~\mu$m were detected \change{with PACS} towards the class I low-mass YSOs as well as the intermediate-mass and class I Herbig Ae sources. \change{No OH emission was detected from the class 0 YSO NGC~1333~IRAS~2A, though the $119~\mu$m was detected in absorption. With HIFI, the 163.12$~\mu$m was not detected from HH~46 and only tentatively detected from NGC~1333~IRAS~2A.}
   The combination of the PACS and HIFI results for HH~46 \change{constrains} the line width (FWHM $\gtrsim$ 11$~$km~s$^{-1}$) and \change{indicates} that the OH emission likely originates from shocked gas. \change{This scenario is supported by trends of the OH flux increasing with the [\ion{O}{i}] flux and the bolometric luminosity, as found in our sample. 
   Similar OH line ratios for most sources suggest that OH has comparable excitation temperatures despite the different physical properties of the sources.}
   }
   {}

\keywords{Astrochemistry --- Stars: formation --- ISM: molecules --- ISM: jets and outflows --- ISM: individual objects: HH~46}

\maketitle

\section{Introduction} 
The hydroxyl radical (OH) is a cornerstone species of the oxygen chemistry in dense clouds and is particularly important in the chemical reaction network of water. H$_2$O and OH are closely linked through the OH + H$_2$ $\Longleftrightarrow$ H$_2$O + H reaction\change{s}. The formation path of H$_2$O from OH is efficient at the high temperatures found in shocks or in the innermost parts of circumstellar envelopes \citep{Kaufman96,Charnley97}. Below about 250~K, standard gas-phase chemistry applies, in which \change{H$_2$O} is formed and destroyed through ion-molecule reactions. In regions not completely shielded from UV radiation, photo-dissociation becomes a major destruction path \change{of H$_2$O, leaving OH as a byproduct}. A better understanding of the OH emission will therefore help \change{to constrain} the water chemistry.

The observation of far-infrared (FIR) rotational OH lines by ground based facilities is severely limited by the Earth's atmosphere. Previous studies of OH FIR emission with the Infrared Space Observatory (ISO) showed that OH is one of the major molecular coolants in star-forming regions \citep[e.g. ][]{Giannini01}. However, with a large beam of 80$\arcsec$, ISO was unable to resolve the central source from the outflows, preventing an assessment of the origin of the OH emission. Interpretation of the ISO OH measurements thus relied mostly on the assumption that the OH emission arises from \change{gas with the same temperature and density} as the \change{high-J} CO \change{FIR} emission \citep[e.g.][]{Nisini99,Ceccarelli98}. The \textit{Herschel Space Observatory} permits observations of OH FIR transitions at both higher angular and spectral resolution and a\change{t} higher sensitivity than ISO. 

Observations of \change{H$_2$O, OH} and related species towards a large set of young stellar objects over a wide range of luminosities and masses are being carried out in the `Water In Star-forming regions with \textit{Herschel}' (WISH) key program to trace the water chemistry during protostellar evolution (van Dishoeck et al. in prep). 
\change{OH emission at 163.12$~\mu$m (1837.8~GHz) was detected with PACS towards the class I YSO HH~46 \citep{vanKempen10}, but the triplet, split by 90~MHz, could not be resolved. Based on modeling results, the OH emission was attributed to a $J$-type shock and not to the quiescent envelope. 
We carried out high-resolution spectroscopy with HIFI to test whether the detected OH emission is dominated by shock contribution by resolving the line profiles. This paper presents the HIFI observations in HH~46 as well as the class 0 object NGC~1333~IRAS~2A. New PACS observations of OH and [\ion{O}{i}] are reported for the low-mass protostars IRAS~15398-3359, NGC~1333~IRAS~2A, and TMR~1.}

\section{Observations and data reduction}
\label{sec:obs}
Because of spin-orbit interaction, the OH rotational levels are built \change{within} two ladders, ${}^2\Pi_{3/2}$ and  ${}^2\Pi_{1/2}$ \citep{Offer92}. Each level is further split by $\Lambda$ doubling and hyperfine structure. A level diagram can be found in Fig. \ref{fig:oh_levels} in the appendix. We use the molecular data from the Leiden atomic and molecular database LAMDA \footnote{http://www.strw.leidenuniv.nl/$\sim$moldata/} \citep{Schoier05}.

High-resolution observations of the OH triplet at \change{163.12$~\mu$m (1837.747, 1837.817 and 1837.837~GHz)} were performed with the Heterodyne Instrument for the Far-Infrared \citep[HIFI,][]{DeGraauw10} on board the ESA \textit{Herschel Space Observatory} (Pilbratt et al. 2010) towards HH~46 and NGC~1333~IRAS~2A. 
HIFI data were stitched together using the Herschel Interactive Processing Environment \citep[HIPE \change{v3.0.1},][]{Ott10} and further analyzed using GILDAS-CLASS\footnote{http://www.iram.fr/IRAMFR/GILDAS} software. We removed standing waves after the subtraction of a low-order polynomial and calibrated to $T_{\mathrm{mb}}$ scale using \change {a main beam efficiency of 0.74}. \change{The H and V polarizations were combined.}

\change{HH~46, TMR~1, IRAS~15398-3359, and NGC~7129~FIRS~2 were observed with PACS (Poglitsch et al. 2010) in line spectroscopy mode around four OH doublets at 79, 84, 119, and $163~\mu$m.  The [\ion{O}{i}] 63 and 145.5$~\mu$m lines were also observed except for TMR~1. Each segment at $\lambda < 100~\mu$m  and $\lambda > 100~\mu$m covers 1 and $~2\mu$m at $R \sim 3000$ and $R \sim 1500$, respectively. In addition, DK~Cha and NGC~1333~IRAS~2A were observed with PACS from 55--210 $\mu$m at $R \sim 1000$ in range spectroscopy mode. Details of the PACS observations of HH~46, NGC~7129~FIRS~2, and DK~Cha are described in \citet{vanKempen10}, \citet{Fich10}, and \citet{vanKempen10digit}, respectively. All spectra were reduced with HIPE v2.9.
PACS spectra are recorded in a $5\times5$ array of $9\farcs4$ square spatial pixels (spaxels). The observations of IRAS~15398-3359, TMR~1, NGC1333~IRAS~2A, and DK~Cha were mispointed sufficiently in a way that the peak of the continuum emission differs from the central spaxel.}

\change{The line and continuum emission seen by PACS is spatially extended in most cases. Spectra were extracted from the spaxels that include OH emission. The wavelength-dependent point-spread function was roughly corrected by comparing the amount of continuum emission in the summed spaxels to that in the total array.}
Most integrated line intensities were derived from Gaussian fits to the unresolved lines. In a few cases, the OH doublets could not be resolved and the intensities were then derived by simple integration over the spectrum.  The absolute flux calibration below and above $100~\mu$m was separately determined from in-flight observations of (point) calibration sources. The relative spectral response function within each band was determined from ground calibration prior to launch. The uncertainty in absolute and relative fluxes is estimated to be 30-50\%. Additional details on the observations can be found in the appendix.

\placefigureHIFISpectrum

\section{Results}
\label{sec:res}
HIFI did not detect the \change{163.12$~\mu$m} OH hyperfine triplet at the noise level of \change{$T_{\mathrm{rms}}\approx 70~$mK} on $T_{\mathrm{mb}}$ scale for the nominal WBS resolution of about 1.1~MHz ($0.163~\mathrm{km}~\mathrm{s}^{-1}$). The spectrum is presented in Fig. \ref{fig:hh46_hifi}.
\change{Combining the HIFI and PACS observations constrains the line width (see Sect. \ref{sec:analysis}).} 

PACS detected OH emission at $79, 84, 119$, and $163~\mu$m towards the class I sources HH~46, TMR~1 and IRAS~15398-3359 as well as the class I Herbig Ae star DK~Cha and the intermediate-mass source NGC~7129~FIRS~2 (Fig. \ref{fig:pacs_spectra}). \change{An} exception is the $163.40~\mu$m line of HH~46, where we only have an upper limit because of the uncertain baseline towards the end of the spectral window where the line is located. The [\ion{O}{i}] $63~\mu$m and $145~\mu$m lines were also \change{detected}. An overview of the results is given in Table \ref{tab:fluxes}. The \change{1~$\sigma$} errors listed in Table \ref{tab:fluxes} do not include the systematic error of 30-50\% from uncertainty in the \change{calibration}.
\change{Because of blending with CO(31-30), the $84.42~\mu$m doublet component of OH is not listed. For the analysis,} we assumed the flux to be identical to the $84.60~\mu$m OH flux \change{because} all observed doublets show comparable fluxes of the components within the calibration uncertainty. 

The only class 0 YSO in the sample, NGC~1333~IRAS~2A, is fundamentally different from all other sources in the sample. The OH $119~\mu$m doublet is seen in absorption, with an equivalent width of about $7.5~\mathrm{km}~\mathrm{s}^{-1}$ for each component. 
No other OH lines were detected at the noise level obtained after removal of the fringing effects. However, \change{the upper limits are larger than the fluxes detected from the class I sources. On the other hand, the upper limit on the [\ion{O}{i}] $63~\mu$m emission is at least a factor of four lower than the weakest [\ion{O}{i}] $63~\mu$m line found in our sample}.

\section{Analysis}
\label{sec:analysis}
\change{To constrain the line width of the $163.12~\mu$m OH triplet from HH~46, the HIFI non-detection is combined with the flux \mbox{($22 \times 10^{-18}~\mathrm{W}~\mathrm{m}^{-2}$)} of the unresolved lines measured with the high-sensitivity PACS instrument.}
The expected integrated intensity for HIFI is $\sim 1.1~\mathrm{K}~\mathrm{km}~\mathrm{s}^{-1}$ \change{for the triplet} in total, as calculated from the PACS observation. \change{This value was derived for a $13\arcsec$ beam under the assumption that the source is unresolved, because the OH emission from HH~46 appears to be centrally concentrated \citep{vanKempen10}.} 
Assuming LTE ratios ($\sim A_{\mathrm{ul}} g_\mathrm{u}$) for the \change{three} components, the strongest transition has an integrated intensity of $0.66~\mathrm{K}~\mathrm{km}~\mathrm{s}^{-1}$. An upper limit on the \change{integrated intensity} is calculated \change{from the HIFI observation} using $\sigma = 1.3 \sqrt{\delta v \Delta v} T_{\mathrm{rms}}$ with \change{$\delta v $ the velocity resolution, $\Delta v$ the expected line width and} an assumed $30~\%$ calibration uncertainty. Rebinning the spectrum to a resolution of 16 MHz ($2.6~\mathrm{km}~\mathrm{s}^{-1}$) yields an rms of \change{$31~$mK} and would therefore allow $5\sigma$ ($3\sigma$) detections for \change{Gaussian lines with $\mathrm{FWHM} \leq 4~\mathrm{km}~\mathrm{s}^{-1}$ ($\mathrm{FWHM} \leq 11~\mathrm{km}~\mathrm{s}^{-1}$).} \change{From the non-detection we conclude that} the flux observed with PACS is likely to be emitted from lines with FWHM $\gtrsim 11~\mathrm{km}~\mathrm{s}^{-1}$. For comparison, the FWHM of the H$_2$O $1_{1 0} - 1_{0 1}$ transition derived from recent unpublished HIFI observations is about \change{$16~\mathrm{km}~\mathrm{s}^{-1}$}. \change{The CO(6-5) and CO(7-6) lines observed with APEX by \citet{vanKempen09hh46} are narrower.}
\placefigurePACSpectrum
\placetableone
\change{In a similar HIFI observation, OH emission was tentatively detected below the 5$\sigma$ level for the strongest triplet component in NGC~1333~IRAS~2A, in agreement with the upper limits on OH emission in the PACS observation of the same source. Given the uncertain baseline, this needs to be treated with caution.}

\placefigureFluxEup

\change{Figure \ref{fig:flux_vs_eup} compares the OH fluxes from YSOs as measured here with PACS and in ISO observations of two additional sources, Ser~SMM1 \citep{Larsson02} and NGC~1333~IRAS~4 \citep{Giannini01}.}
Despite the wide range of luminosities and masses covered in our sample, the sources show surprisingly similar characteristics in terms of their OH emission: \change{t}he OH $84~\mu$m doublet is generally the strongest of the four, while the integrated intensity of the weakest doublet at $163~\mu$m is roughly a factor of three lower. \change{
The $119~\mu$m flux can vary compared to higher excitation lines because this line, which is connected to the ground state level, becomes more easily optically thick. It can also be affected by absorption against the continuum by cold gas layers lying in front of the source.} 
The $79~\mu$m doublet links the lowest energy states of both rotational ladders, but because of the smaller Einstein $A$ coefficients of the cross-ladder transitions, these lines are less affected by optical depth than the $119~\mu$m lines. The \change{$79~\mu$m fluxes} are lower than \change{those} at $84~\mu$m for most sources, but higher than \change{at} $119~\mu$m.

Flux ratios of the observed OH lines were calculated \change{from the summed doublet components} to compare the excitation of the involved energy levels among the sources. The line ratios of different sources agree within a factor of two except when the 119$~\mu$m transition is involved (Fig. \ref{fig:oh_ratios} in the appendix) \change{and are consistent with the results from ISO observations from the class~0 sources NGC~1333~IRAS~4 and Ser~SMM1}. 

\change{Emission in the $84~\mu$m transition, which has $E_{\mathrm{up}}/k_{\mathrm{B}} \approx 300~\mathrm{K}$, indicates that OH is tracing the  warm ($T \gtrsim 100~\mathrm{K}$) and dense ($n \gtrsim 10^5~\mathrm{cm}^{-3}$) gas in our sources. Modeling results show that transitions in the ${}^2\Pi_{3/2}$ ladder are mostly excited by collisions while the population of the ${}^2\Pi_{1/2}$ levels is dominated by radiative pumping via the cross-ladder transitions. The weak $163~\mu$m lines and \change{emission at $79~\mu$m indicate} that FIR pumping is less important than collisional excitation in our sources.}

\change{Figure \ref{fig:flux_lum} shows the trend of stronger OH emission with increasing bolometric source luminosity $L_{\mathrm{bol}}$ found in our source sample. We calculated the OH line luminosity $L_{\mathrm{OH}}$ individually for each transition and source from the observed fluxes} and found indications that the differences \change{between} the sources \change{may} depend on the individual $L_{\mathrm{bol}}$. The correlation between $L_{\mathrm{OH}}$ and $L_{\mathrm{bol}}$ is reminiscent of that found for the CO outflow force with luminosity \citep{Bontemps96}. The latter relation was taken as evidence that more massive envelopes have higher accretion rates and thus drive more powerful outflows. \citet{vanKempen10} \change{speculate} that the OH emission originates in the wake of the jet impinging on the dense, inner parts of the envelope, \change{creating dissociative shocks in which} [\ion{O}{i}] is the dominant coolant, followed by OH \citep{Neufeld89}. The relation between OH emission and luminosity \change{supports} this scenario.
\placefigureFluxLum

\placefigureFluxOxygen

\change{Comparison of OH with [\ion{O}{i}] emission shows that stronger OH emission coincides with higher [\ion{O}{i}] intensities \change{(Fig. \ref{fig:flux_ox})} and also with increasing [\ion{O}{i}] 63/145$~\mu$m line flux ratios.}
For HH~46, TMR~1, and NGC~7129~FIRS~2, the bulk of the [\ion{O}{i}] and OH emission comes from close to the protostar where densities are on the order of $10^5~\mathrm{cm}^{-3}$ or higher, as illustrated by the lack of extended OH emission in HH~46 \citep{vanKempen10}. Some spatially extended OH emission is detected from IRAS~15398-3359 and DK~Cha, and is highly correlated with the spatial extent of [\ion{O}{i}] emission.  The correlation between the intensities \mbox{(Fig. \ref{fig:flux_ox})} suggests that the bulk of [\ion{O}{i}] and OH emission originates in the same physical component in all sources. \change{This argument, together with the \mbox{OH -- $L_{\mathrm{bol}}$} and \mbox{OH -- [\ion{O}{i}]} relations, supports the dissociative shock scenario.} The [\ion{O}{i}] \mbox{63$\mu$m / 145$\mu$m} line ratios are in the range of 13--19, \change{also} consistent with fast, dissociative $J$-type shocks \citep[\mbox{$\varv >$ 60 km s$^{-1}$},][]{Neufeld89}. \change{Note that an \mbox{OH -- [\ion{O}{i}]} relation can also be indicative of photo-dissociation, as argued for Sgr~B2 by \citet{Goicoechea04}. Models of photon-dominated regions \citep{Kaufman99} predict similar [\ion{O}{i}] \mbox{63$\mu$m / 145$\mu$m} line ratios, but those require $n < 10^5~\mathrm{cm}^{-3}$, which is inconsistent with an origin in the inner, dense envelope.}

\section{Conclusions}
The OH hyperfine transition triplet at \change{163.12$~\mu$m (1837.8~GHz)} was not detected above the noise level obtained with HIFI. Combined with the flux derived from the unresolved line observed with PACS, this constrains the line width to \mbox{FWHM $\gtrsim 11~\mathrm{km}~\mathrm{s}^{-1}$}. This width is much broader than expected for the \change{quiescent} envelope from ground-based observations \citep{vanKempen09hh46} and indicates that the observed OH emission most likely stems from shocked gas in HH~46.
 
\textit{Herschel} PACS observations of OH lines at $79, 84, 119$, and $163~\mu$m have been carried out for four low-mass YSOs, one intermediate-mass protostar and one class I Herbig Ae object. OH emission is detected in all sources except the class 0 YSO NGC~1333~IRAS~2A, where the OH 119$~\mu$m transitions are observed in absorption and only upper limits can be derived for the other lines. Sources with detected OH emission show surprisingly similar OH line ratios despite the large ranges of physical properties covered in this study, suggesting that OH emission might arise from gas at similar conditions in all sources. Furthermore, we find trends of correlations between OH integrated intensities and [\ion{O}{i}] emission as well as \change{bolometric luminosity, consistent with an origin in the wake of dissociative shocks.} Given the low number of sources in the sample, confirmation from additional observations \change{is needed}.

Further OH observations \change{and modeling should} eventually allow the determination of the OH/H$_2$O abundance ratio in shocks, which traces the UV field through its dependence on the fraction of atomic to molecular hydrogen.

\bibliographystyle{aa}
\bibliography{mybib}

\begin{thebibliography}{18}
\expandafter\ifx\csname natexlab\endcsname\relax\def\natexlab#1{#1}\fi

\bibitem[{{Bontemps} {et~al.}(1996){Bontemps}, {Andre}, {Terebey}, \&
  {Cabrit}}]{Bontemps96}
{Bontemps}, S., {Andre}, P., {Terebey}, S., \& {Cabrit}, S. 1996, \aap, 311,
  858

\bibitem[{{Ceccarelli} {et~al.}(1998){Ceccarelli}, {Caux}, {White}, {Molinari},
  {Furniss}, {Liseau}, {Nisini}, {Saraceno}, {Spinoglio}, \&
  {Wolfire}}]{Ceccarelli98}
{Ceccarelli}, C., {Caux}, E., {White}, G.~J., {et~al.} 1998, \aap, 331, 372

\bibitem[{{Charnley}(1997)}]{Charnley97}
{Charnley}, S.~B. 1997, \apj, 481, 396

\bibitem[{{De Graauw} {et~al.}(2010){De Graauw}, {Helmich}, {Philipps},
  {Stutzki}, {Caux}, \& ...}]{DeGraauw10}
{De Graauw}, T., {Helmich}, F.~P., {Philipps}, T.~G., {et~al.} 2010, submitted
  to this issue

\bibitem[{{Fich} {et~al.}(2010){Fich}, {Johnstone}, {van Kempen}, {M$^{\textrm
  c}$Coey}, {Fuente}, {Caselli}, {Kristensen}, \& {Plume}}]{Fich10}
{Fich}, M., {Johnstone}, D., {van Kempen}, T.~A., {et~al.} 2010, \aap ~in press

\bibitem[{{Giannini} {et~al.}(2001){Giannini}, {Nisini}, \&
  {Lorenzetti}}]{Giannini01}
{Giannini}, T., {Nisini}, B., \& {Lorenzetti}, D. 2001, \apj, 555, 40

\bibitem[{{Goicoechea} {et~al.}(2004){Goicoechea},
  {Rodr{\'{\i}}guez-Fern{\'a}ndez}, \& {Cernicharo}}]{Goicoechea04}
{Goicoechea}, J.~R., {Rodr{\'{\i}}guez-Fern{\'a}ndez}, N.~J., \& {Cernicharo},
  J. 2004, \apj, 600, 214

\bibitem[{{Kaufman} \& {Neufeld}(1996)}]{Kaufman96}
{Kaufman}, M.~J. \& {Neufeld}, D.~A. 1996, \apj, 456, 611

\bibitem[{{Kaufman} {et~al.}(1999){Kaufman}, {Wolfire}, {Hollenbach}, \&
  {Luhman}}]{Kaufman99}
{Kaufman}, M.~J., {Wolfire}, M.~G., {Hollenbach}, D.~J., \& {Luhman}, M.~L.
  1999, \apj, 527, 795

\bibitem[{{Larsson} {et~al.}(2002){Larsson}, {Liseau}, \&
  {Men'shchikov}}]{Larsson02}
{Larsson}, B., {Liseau}, R., \& {Men'shchikov}, A.~B. 2002, \aap, 386, 1055

\bibitem[{{Neufeld} \& {Dalgarno}(1989)}]{Neufeld89}
{Neufeld}, D.~A. \& {Dalgarno}, A. 1989, \apj, 344, 251

\bibitem[{{Nisini} {et~al.}(1999){Nisini}, {Benedettini}, {Giannini}, {Caux},
  {di Giorgio}, {Liseau}, {Lorenzetti}, {Molinari}, {Saraceno}, {Smith},
  {Spinoglio}, \& {White}}]{Nisini99}
{Nisini}, B., {Benedettini}, M., {Giannini}, T., {et~al.} 1999, \aap, 350, 529

\bibitem[{{Offer} \& {van Dishoeck}(1992)}]{Offer92}
{Offer}, A.~R. \& {van Dishoeck}, E.~F. 1992, \mnras, 257, 377

\bibitem[{{Ott}(2010)}]{Ott10}
{Ott}, S. 2010, in ASP Conference Series, Astronomical Data Analysis Software
  and Systems XIX, Y. Mizumoto, K.-I. Morita, and M. Ohishi, eds., in press.

\bibitem[{{Sch{\"o}ier} {et~al.}(2005){Sch{\"o}ier}, {van der Tak}, {van
  Dishoeck}, \& {Black}}]{Schoier05}
{Sch{\"o}ier}, F.~L., {van der Tak}, F.~F.~S., {van Dishoeck}, E.~F., \&
  {Black}, J.~H. 2005, \aap, 432, 369

\bibitem[{{van Kempen} {et~al.}(2010{\natexlab{a}}){van Kempen}, {Green},
  {Evans}, {van Dishoeck}, {Kristensen}, {Herczeg}, {Merin}, {Lee},
  {Joergensen}, {Bouwman}, {Acke}, {Adamkovics}, {Augereau}, {Bergin}, {Blake},
  {Brown}, {Carr}, {Chen}, {Cieza}, {Dominik}, {Dullemond}, {Dunham},
  {Glassgold}, {G{\"u}del}, {Harvey}, {Henning}, {Hogerheijde}, {Jaffe}, {Kim},
  {Knez}, {Lacy}, {Maret}, {Meeus}, {Meijerink}, {Mulders}, {Mundy}, {Najita},
  {Olofsson}, {Pontoppidan}, {Salyk}, {Sturm}, {Visser}, {Waters}, {Waelkens},
  \& {Y{\i}ld{\i}z}}]{vanKempen10digit}
{van Kempen}, T.~A., {Green}, J.~D., {Evans}, N.~J., {et~al.}
  2010{\natexlab{a}}, \aap ~in press

\bibitem[{{van Kempen} {et~al.}(2010{\natexlab{b}}){van Kempen}, {Kristensen},
  {Herczeg}, {Visser}, {van Dishoeck}, {Wampfler}, {Bruderer}, {Benz}, {Doty},
  {Brinch}, {Hogerheijde}, {J{\o}rgensen}, {Tafalla}, {Neufeld}, {Bachiller},
  {Baudry}, {Benedettini}, {Bergin}, {Bjerkeli}, {Blake}, {Bontemps}, {Braine},
  {Caselli}, {Cernicharo}, {Codella}, {Daniel}, {di Giorgio}, {Dominik},
  {Encrenaz}, {Fich}, {Fuente}, {Giannini}, {Goicoechea}, {de Graauw},
  {Helmich}, {Herpin}, {Jacq}, {Johnstone}, {Kaufman}, {Larsson}, {Lis},
  {Liseau}, {Marseille}, {McCoey}, {Melnick}, {Nisini}, {Olberg}, {Parise},
  {Pearson}, {Plume}, {Risacher}, {Santiago-Garcia}, {Saraceno}, {Shipman},
  {van der Tak}, {Wyrowski}, {Yildiz}, {Ciechanowicz}, {Dubbeldam}, {Glenz},
  {Huisman}, {Lin}, {Morris}, {Murphy}, \& {Trappe}}]{vanKempen10}
{van Kempen}, T.~A., {Kristensen}, L.~E., {Herczeg}, G.~J., {et~al.}
  2010{\natexlab{b}}, \aap ~in press

\bibitem[{{van Kempen} {et~al.}(2009){van Kempen}, {van Dishoeck},
  {G{\"u}sten}, {Kristensen}, {Schilke}, {Hogerheijde}, {Boland}, {Nefs},
  {Menten}, {Baryshev}, \& {Wyrowski}}]{vanKempen09hh46}
{van Kempen}, T.~A., {van Dishoeck}, E.~F., {G{\"u}sten}, R., {et~al.} 2009,
  \aap, 501, 633

\end{thebibliography}

\begin{acknowledgements}
The work on star formation at ETH Zurich is partially funded by the Swiss National Science Foundation (grant nr. 200020-113556). This program is made possible thanks to the Swiss HIFI guaranteed time program.
HIFI has been designed and built by a consortium of institutes and university departments from across Europe, Canada and the United States under the leadership of SRON Netherlands Institute for Space
Research, Groningen, The Netherlands and with major contributions from Germany, France and the US.
Consortium members are: Canada: CSA, U.Waterloo; France: CESR, LAB, LERMA, IRAM; Germany:
KOSMA, MPIfR, MPS; Ireland, NUI Maynooth; Italy: ASI, IFSI-INAF, Osservatorio Astrofisico di Arcetri-
INAF; Netherlands: SRON, TUD; Poland: CAMK, CBK; Spain: Observatorio Astronómico Nacional (IGN),
Centro de Astrobiología (CSIC-INTA). Sweden: Chalmers University of Technology - MC2, RSS \& GARD;
Onsala Space Observatory; Swedish National Space Board, Stockholm University - Stockholm Observatory;
Switzerland: ETH Zurich, FHNW; USA: Caltech, JPL, NHSC.
\end{acknowledgements}

\Online

\begin{appendix}

\section{Observational details}
\placetabletwo
\change{Table \ref{tab:source_props} lists the coordinates, observing dates, and the observation identity numbers of our sources along with the assumed distance, the bolometric luminosity, the envelope mass and the source type. For comparison, Ser~SMM1 and NGC~1333~IRAS~4 observed with ISO are included as well. For Ser~SMM1, we use the average values of the fluxes presented by \citet{Larsson02}. The data for NGC~1333~IRAS~4 are taken from \citet{Giannini01}. Note that for NGC~1333~IRAS~4, we use the luminosity and mass of NGC~1333~IRAS~4A.}

\section{OH term diagram}
\placefigureOHladder

\section{OH line ratios}
\placefigureRatios

\end{appendix}

\end{document}